# In situ measurements of the variable slow solar wind near sector boundaries

E. Sanchez-Diaz[1], A. P. Rouillard[1], B. Lavraud[1], E. Kilpua[2], J. A. Davies[3]

[1] *Institut de Recherche en Astrophysique et Planétologie, University of Toulouse, CNRS, UPS, CNES, 9 avenue Colonel Roche, BP 44346-31028, Toulouse Cedex 4A, France;* eduardo.sanchez-diaz@irap.omp.eu, alexis.rouillard@irap.omp.eu, benoit.lavraud@irap.omp.eu

[2]*Space Physics Department, Department of Physics, P.O. Box 64 FI-00014, University of Helsinki, Helsinki, Finland;* emilia.kilpua@helsinki.fi

[3] *RAL Space, STFC-Rutherford Appleton Laboratory, Harwell Campus, Didcot OX11 0QX, UK;* jackie.davies@stfc.ac.uk

## Abstract

The release of density structures at the tip of the coronal helmet streamers, likely as a consequence of magnetic reconnection, contributes to the mass flux of the slow solar wind. In situ measurements in the vicinity of the heliospheric plasma sheet of the magnetic field, protons and suprathermal electrons reveal details of the processes at play during the formation of density structures near the Sun. In a previous article, we exploited remote-sensing observations to derive a 3-D picture of the dynamic evolution of a streamer. We found evidence of the recurrent and continual release of dense blobs from the tip of the streamers. In the present paper, we interpret in situ measurements of the slow solar wind during solar maximum. Through both case and statistical analysis, we show that in situ signatures (magnetic field magnitude, smoothness and rotation, proton density and suprathermal electrons, in the first place) are consistent with the helmet streamers producing, in alternation, high-density regions (mostly disconnected) separated by magnetic flux ropes (mostly connected to the Sun). This sequence of emission of dense blobs and flux ropes also seems repeated at smaller scales inside each of the high-density regions. These properties are further confirmed with in situ measurements much closer to the Sun using Helios observations. We conclude on a model for the formation of dense blobs and flux ropes that explains both the in situ measurements and the remote-sensing observations presented in our previous studies.

## 1. Introduction

While the fast solar wind is thought to form along the open magnetic field lines rooted near the center of large coronal holes (e.g. Hollweg & Isenberg 2002), the origin of the Slow Solar Wind (SSW) is still the subject of much debate and represents a major gap in our understanding of the heliosphere. The frequent release of density enhancements with radial dimensions of several solar radii ($R_s$), usually called "blobs", is considered an important source of the SSW. They emerge continually from coronal streamers and accelerate from less than 100 km/s at 2 $R_s$, up to 300 km/s at 30 $R_s$, the latter being the typical speed of the SSW in the heliosphere (e.g., Sheeley et al. 1997). Sanchez-Diaz et al. (2017a), hereafter SD17A, found a radial extent of 12 $R_s$ for these small transients when they were observed 30 $R_s$ above the center of the Sun with the Solar Terretrial Relationships Observatory (STEREO; Kaiser et al. 2008) Helisopheric Imagers (HI; Eyles et al. 2009). SD17A also found that this release of small transients occurs with a periodicity 19.5 hours. According to these sizes and periodicity of release, they concluded that the release of such blobs could account for 15% of the SSW during solar maximum conditions. This is consistent with blobs being one of the two main sources of SSW according to the works of Kasper et al. (2007, 2012), McGregor (2011) and Stakhiv et al. (2015, 2016).

Coronagraph images have revealed that even smaller density structures, with radial extents of the order of 1 $R_s$, known as Periodic Density Structures (PDSs), also emerge from coronal streamers, at a rate of one every 60 to 100 minutes (Viall et al. 2010; Viall & Vourlidas 2015; Di Matteo et al. 2019). PDSs are speculated to be associated with the substructure of blobs (Viall & Vourlidas 2015). The advent of high-cadence, in situ composition measurements has revealed the presence of similar quasi-periodic variations in the abundance of minor ions in the SSW (Kepko et al. 2016). The latter could relate to the PDSs observed in coronal images (Viall et al. 2010; Viall & Vourlidas 2015). According to in situ measurements, between 70 and 80% of the SSW could be made up of such quasi-periodic structures (Viall et al. 2008). In this paper, we exploit in situ measurements to study the substructure of blobs and investigate their possible relation to PDSs.

Blobs have mostly been observed in running-difference images (e. g. Sheeley et al. 1997, 2009), which are generated by subtracting, from each image, the previous one. In such images, blobs manifest as brightness variations that propagate radially outwards. Using coronagraph running-difference images from the early years of the STEREO mission, Sheeley et al. (2009) presented evidence that blobs are the signatures of small-scale expanding coronal loops. By comparing similar remote-sensing imagery to in situ measurements at 1 AU, Rouillard et al. (2010ab) found that blobs measured in situ near 1 AU correspond to density enhancements near the back-end of magnetic structures that have the typical properties of magnetic flux ropes. This suggests that the expanding bright loops observed by Sheeley et al. (2009) could also be associated with dense plasma accumulated at the back end of magnetic flux ropes. The association of blobs with the release of flux ropes is also supported by the inward motion of plasma in the solar corona that is systematically associated with the outward release of large blobs (Sanchez-Diaz et al. 2017b, hereafter SD17B). This fact was interpreted as evidence that magnetic reconnection, taking place at around 5-6 $R_s$ from Sun center, is the mechanism by which blobs are released. SD17A detected a blob in situ and found that the flux rope, in that case, preceeded the density enhancement rather than being concomitant with it. The fact that dense blobs and flux ropes come in alternation as a result of the release process is also adressed in the present paper.

The direct comparison of remote-sensing observations of blobs with their in situ counterparts has only been possible under particular orbital configurations. This limitation is determined by (1) the observational constraints imposed by Thomson scattering, (2) the rapid expansion of the solar wind and (3) the associated decrease of solar wind density with radial distance. Despite the long exposure times of the HI instruments on board STEREO, blobs expanding in the interplanetary medium disappear rapidly as they propagate through the HI-1 field of view. To be observed further out, this density fall off needs to be countered by compression at a Stream Interaction Region, SIR (called a Corotating Interaction Region, CIR, if it survives for more than one solar rotation; Jian et al. 2006). Moreover, when the line of sight from the observer is aligned along the tangent of the dense spiral formed by the CIR, compressed blobs can be tracked all the way out to 1 AU (Rouillard et al. 2009, Sheeley & Rouillard 2010). Such favorable conditions were met from STEREO-B's perspective during the early years of the mission, while both spacecraft were close to each other and to the Earth and might be met by STEREO-A when it approaches Earth again. Blobs emitted from a source longitude located 17° westward of STEREO-B are very clearly observed in the J-maps out to distances of 1 AU due to the favourable alignment with the Thomson geometry (Rouillard et al. 2009). Therefore, when a spacecraft taking in situ data is located about 17° eastward of STEREO-B, it is possible to carry out simultaneous remote-sensing and in situ measurements of the same blob. Rouillard et al. (2010a,b) exploited such an orbital configuration during the early years of the STEREO mission to track blobs from the Sun to ACE and STEREO-A. In situ signatures of blobs similar to those presented by Rouillard et al. (2010b) have been reported for blobs that lie outside CIRs (e. g. Kilpua et al. 2009). Apart from the interval studied by Rouillard et al. (2010a,b), it has not been possible to compare simultaneous remote-sensing observations and in situ measurements of blobs.

After 2010, when the Earth and the STEREO spacecraft were more than 70° apart in longitude, the direction of motion of an individual blob estimated from HI imagery was no longer sufficiently accurate to ascertain whether an individual blob would impact the Earth or the other STEREO spacecraft. Our approach in this paper is, therefore, to exploit the results of the imagery-based analysis in SD17A in order to set the context for the analysis of in situ data. Instead of tracking individual blobs all the way out to 1 AU, as was done in Rouillard et al. (2010ab), we analyse the in situ data to derive the trajectory of the spacecraft through the streamer structures inferred by SD17A from white-light images. Note that, contrary to previous analysis (e. g. Rouillard et al. 2010ab, Plotnikov et al. 2016), the method used in this paper is not restricted to blobs entrained by CIRs and provides the basis for the interpretation of in situ data during the passage of blobs that cannot be tracked to 1 AU using heliospheric imaging.

## *1.1. Interpretation of in situ measurements in terms of the spatial distribution of blobs and flux ropes in the heliosphere*

Figure 1, repeated from SD17A, presents a schematic of the size and spatial distribution of transient structures released by a highly-tilted neutral line encountered during elevated solar activity. In this picture, deduced from the analysis of STEREO and SoHO imagery, the Heliospheric Current Sheet (HCS) forms a spiral structure filled with High Density Regions (HDRs) and intervening lower density regions, observed respectively as bright and dark regions in background-subtracted images. It was concluded by SD17AB that these lower density regions must correspond to the locus of loops and twisted magnetic field structures ejected by the streamer and the HDRs form where these loops and twisted magnetic fields reconnect to become flux ropes that may be partly disconnected from the corona. HDRs in the vicinity of the heliospheric plasma sheet (HPS), measured in situ, would therefore correspond to bright blobs imaged in coronagraphs, should therefore form at the backend of ejected flux ropes.

In the present paper, we use in situ measurements to evaluate this hypothesis. Hereafter, we will refer to the HDRs as the in situ equivalent to "blobs", and we will refer to the space between consecutive blobs as flux ropes. Figure 1a shows a latitudinal cut (i.e. in a plane of constant heliographic longitude) through the centre of a blob. We represent the blob as an ellipse with major and minor axes of dimension 12 $R_s$ and 5 $R_s$, respectively. SD17A's results suggest that blobs are typically separated by 15° from one another (in latitude in this case of a highly-tilted current sheet) and have an extent along the current sheet (latitudinal extent in this case) of 5 $R_s$ when they reach a height of 30 $R_s$. SD17A also found that blobs are released roughly every 20 h, simultaneously all along the neutral line, forming (in the case of a highly-tilted neutral line) latitudinally extended and alternating bands of increased density and low-density flux ropes propagating outwards in the heliosphere. Figure 1b shows, for a fixed time, the spatial distribution of small-scale transients along the Parker spiral in a plane of constant heliographic latitude. In Figure 1b, blobs (or HDRs) are marked with gray shading (as in Figure 1a) and the space in between consecutive blobs (i.e. the intervening flux ropes) with dashed ellipses. Figure 1c shows the distribution of blobs in a plane of constant heliographic longitude. The spatial distribution of small transients depicted in Figure 1 is used to interpret, in a unified manner, the in situ signatures of small transients measured in the SSW near the HCS as well as the signatures of the peculiar periods of very slow solar wind measured at Helios and analysed by Sanchez-Diaz et al. (2016). As we stand at the dawn of a golden age in heliophysics, these results can be used to predict the expected in situ signatures of the SSW that will be measured by Solar Orbiter and Parker Solar Probe at varying heliospheric distances from the Sun.

Based on Figure 1b, which displays the distribution of blobs and flux ropes for the case of a north-south oriented neutral line, we can consider several possible spacecraft trajectories through the sector boundary crossing. We can distinguish the following scenarios, which are sketched with black dashed arrows in Figure 1b:

> **Scenario 1 (HDR + flux rope):** The spacecraft crosses the sector boundary by first intersecting a blob and then a flux rope. A spacecraft intersecting the edge of the flux rope would only measure the poloidal magnetic field component of the rope. If the central axis of the flux rope was vertical, the spacecraft would measure the rotation of the magnetic field primarily in the equatorial plane. The spacecraft intersection of the HCS would manifest as a HDR with highly variable plasma and magnetic fields. The magnetic fields inside the HDR would likely be lower as a result of magnetic field annihilation, due to reconnection occurring closer in.

**Scenario 2 (HCS):** The spacecraft crosses a sector boundary characterized by the detection of a single, sharp current sheet (the HCS). Such a 'clean' HCS crossing occurs when the neutral line has had time to reform, and thin out, prior to the release of a new flux rope from the tip of the helmet streamer. The expected signature is that of a (nearly) concomitant inversion of the magnetic field polarity and a sector boundary, the latter marked by a clear reversal in the pitch angle distribution of suprathermal electrons. The region immediately surrounding the HCS could be observed as a narrow peak of elevated densities situated at the polarity inversion, typically referred to as the HPS.

**Scenario 3 (HDR):** In this scenario, the spacecraft crosses the blobs. The signature would be a broader HDR lasting up to 10 hours, likely correlated with weaker but complex magnetic fields.

**Scenario 4 (flux rope + HDR):** The reverse trajectory to that of scenario 1. The spacecraft crosses the edge of the magnetic flux rope and the edge of a blob. In this scenario, the in situ measurements would exhibit first the signature of a flux rope crossed near the edge and then that of a HDR.

**Scenario 5 (flux rope):** The spacecraft crosses a magnetic flux rope replacing the HCS without crossing a HDR. Note that the whole streamer, at the tip of which HDRs and flux ropes are forming, is itself a dense structure but image processing tends to subtract the relatively stable background brightness of the streamer in order to reveal the blobs. Therefore, even flux ropes could exhibit relatively high densities but, according to the results of SD17A, necessarily lower than those measured in the HDR and its associated blobs.

In section 2, we will show five examples of in situ measurements of highly-tilted HCS crossings that illustrate all these scenarios. In section 3, we will present the results of a full survey of in situ measurements of highly-tilted HCSs during solar cycle 24. Taking advantage of the 3D space-time distribution of blobs in the heliosphere found by SD17A and according to the 5 different scenarios of HCS crossing described above, this list of events enables us to ascertain a number of important features about blobs and flux ropes, in particular regarding their structure and their origin. Finally, in section 4, we show long-lasting HCS crossings measured by the Helios spacecraft (Sanchez-Diaz et al. 2016) to further test the model presented in Figure 1 with measurements taken much closer to the Sun.

## *2. In situ measurements of 5 different types of HCS crossings at 1 AU*

During solar cycle 24, several crossings of a north-south oriented HCS were measured at 1 AU by the In Situ Measurements of Particles and CME Transients (IMPACT; Luhmann et al. 2008) instrument package on STEREO (Kaiser et al. 2008), which includes the Solar Wind Electron Analyzer (SWEA; Sauvaud et al. 2008) that measures strahl electrons, and by the Solar Wind Experiment (SWE; Olgivie et al. 1995) and Magnetic Field Investigation (MFI; Lepping et al. 1995) instruments, onboard the Wind spacecraft (Harten & Clark 1995). From these HCS crossings, we first extract examples of each of the five different scenarios described above and then undertake a systematic analysis. The five examples shown in Figure 2-6 are all traced back to highly-tilted current sheets at the Sun, such as the one studied by SD17A and SD17B. The link with a highly-tilted HCS is established by exploiting the same Potential Field Source Surface (PFSS) technique (Schatten et al. 1969, Hoeksema et al. 1984, Wang & Sheeley 1992) as employed by Sanchez-Diaz et al. (2016) to track Helios data back to the source region.

**Scenario 1: HDR + flux rope.** Figure 2 presents STEREO-B in situ measurements, from 20:00 UT on 2011 May 19 to 20:00 UT on May 20. The radial component of the magnetic field (panel c) reverses at 05:45 UT on May 20, at the same time as the pitch angle of suprathermal strahl electrons reverses from 0° to 180° (panels a and b). This marks a clear change of magnetic sector (e. g. Crooker et al. 2004; Owens et al. 2014). Note that there is no significant difference in the solar wind speed before and after the sector crossing (panel e). This implies that no CIR-associated compression of the solar wind should be detected adjacent to this crossing. Any HDR detected during this period is therefore of coronal rather than interplanetary origin. At 02:30 UT on May 20 (first vertical black dashed line), the spacecraft enters a HDR with reduced magnetic field. At 04:30 UT (second dashed black line), STEREO-B exits this HDR and enters a region characterized by stronger magnetic fields (about 10 nT) with a smooth variation of the magnetic field components. The latter corresponds to the classic rotation of magnetic field components measured in magnetic flux ropes that here lasts 8 hours.

We interpret this sequence of a 2 h-long HDR followed by a magnetic flux rope passage as the likely signature of Scenario 1. Passage through the HDR (or blob) is coincident with a decrease in the flux of suprathermal electrons in both the parallel and anti-parallel directions. The disappearance of the strahl electrons suggests that the HDR is magnetically disconnected from the Sun. Later, we will evaluate more systematically whether this signature is typical of blobs.

**Scenario 2: HCS.** Figure 3 presents a case of Scenario 2, i.e. the crossing by the Wind spacecraft of a HCS undisturbed by the passage of a transient, that took place in April 2011. Simultaneous reversals of the radial magnetic field component and the pitch angle of the strahl occur at 14:00 UT on April 28 (marked by a vertical black solid line). These are classic signatures of a HCS crossing (e.g. Crooker et al. 2004; Owens et al. 2014). The proton density increases at a nearly constant rate from 03:00 UT to 14:00 UT on April 28. A sharp decrease in density is observed immediately after the magnetic field reversal. Note that, again, there is no significant difference in the solar wind speed before and after the HCS crossing. This lack of localized compression indicates that the elevated densities are of coronal origin rather than being the product of compression ahead of a CIR.

**Scenario 3: blob only.** Figure 4 presents in situ measurements taken by STEREO-A from 2011 June 27 to 29. Similar to the previous cases, a reversal in the sign of the radial component of the magnetic field occurs at the same time as a reversal in the direction of the pitch angle of suprathermal strahl electrons, at about 03:45 UT on June 28. This indicates a HCS crossing (first black dasked line). A HDR, detected between 03:45 and 12:00 UT on June 28, is delimited by the two dashed lines. We associate the 8-hr duration of this HDR to the passage of an entire blob across the spacecraft, matching the duration of blobs observed in HI as demonstrated by SD17A. As in Figure 2, this HDR is associated with a weak interplanetary magnetic field. The correlated occurrence of a HDR and weaker magnetic fields could result from the pressure equilibrium sought by the corona following the reconnection process that released the density structure. We suggest that the whole structure, detected between 03:45 and 12:00 UT, is the in situ counterpart of a blob (as seen in images). These measurements correspond, thus, to Scenario 3 (blob only). This blob is also associated with a suprathermal electron strahl drop-out (i.e., disappearance of the strahl). This drop-out appears more pronounced than in the case for the blob shown in Figure 2 (Scenario 1: HDR and flux rope) because the flux of suprathermal electrons decreases by several orders of magnitude with respect to that of the ambient solar wind. However, in both cases, the fluxes are similar to that of the halo electrons (in the direction opposite to the strahl) either side of the sector crossing. This is consistent with the idea that the blobs are magnetically disconnected from the Sun, and that the observed suprathermal electrons (with low fluxes) are in fact halo electrons that have been backscattered from a location further out in the heliosphere (e.g. Gosling et al. 2005; Lavraud et al., 2010).

Figure 4 reveals that this HDR (blob) is made up of three different density enhancements separated from each other by about 3 hours. The first peak extends from 03:30 to 05:45 UT on June 28, the second one from 06:37 to 08:07 UT and the third one from 09:10 to 10:30 UT. There seems to be a general anti-correlation between density and magnetic field strength inside the blob. The geometry of the Parker spiral at 1 AU, combined with the radial outflow of the solar wind, do not allow two or more different blobs of the sizes inferred by SD17A to be measured during a single sector crossing. This is also because the sector boundary is, in this instance, oriented in a north-south direction (SD17A). Therefore, these smaller-scale features are likely to be substructures of a single blob. The periodicities and sizes of these substructures fall in the range of PDSs described by Viall et al. (2008, 2009, 2010), Viall & Vourlidas (2015), Kepko et al. (2016) and Di Matteo et al. (2019).

**Scenario 4: flux rope + HDR.** Figure 5, repeated from SD17A presents STEREO-A in situ measurements from 11:00 UT on 2013 July 08 to 04:00 UT on July 10. This time interval corresponds to Carrington rotation 2138, exactly one Carrington rotation after the remote-sensing observations of the highly-tilted streamer analyzed by SD17A.

From 18:00 UT on 2013 July 08 to 02:00 UT on July 09, the spacecraft encounters an enhanced (~8 nT) and smooth magnetic field (panel b) consistent with a flux rope. This flux rope is followed by a HDR that lasts for 4 hours, until 06:00 UT, and is concomitant with a weak magnetic field, similar to those in Figures 2 and 4. The period between 02:00 UT and 06:00 UT thus corresponds to the passage of a blob. Figure 5 shows an example of an HCS crossing as depicted in Scenario 4 (the opposite scenario to that demonstrated in Figure 2). The sector boundary is located near the end of the HDR passage, at 05:45UT. As for the two previous cases (shown in Figures 2 and 4), part of the HDR is coincident with a drop out of the strahl electrons, lasting for 2 hours, from 02:00 to 04:00UT. Just as in Figure 2, the whole small transient is associated with a lower temperature than that of the surrounding solar wind (panel f).

The blob passage shown in Figure 5 lasts for 4 hours and contains two separate density peaks (panel d). Because the flux rope is also detected, it is very likely that this crossing samples only the edge of a blob rather than crossing its full extent. Therefore, it is possible that the two peaks are part of a longer-lasting HDR with a substructure similar to that of the blob seen in Figure 4. We note as well that the speed of 400 to 450 km/s around the sector boundary (panel e) is very similar to the speed of blobs emitted by the HCS during the previous Carrington rotation, deduced from HI observations by SD17A.

**Scenario 5: flux rope only.** Figure 6 presents in situ measurements made by the Wind spacecraft from 16:00 UT on 2010 January 30 to 20:00 UT on January 31. As in previous cases, a reversal in the radial component of the magnetic field concomitant with a reversal in strahl pitch angle, at 01:30 UT on January 31, is a clear signature of the passage of the HCS over the spacecraft. This is followed by an enhanced magnetic field exhibiting a flux rope structure. This flux rope passage lasts for 6.5 hrs, from 01:30 to 08:00 UT. In this case, there is no HDR associated with the crossing. Instead, the flux rope replaced the HDR that is often measured in association with a HCS crossing. This is thus an example of Scenario 5 (flux rope only), as described in Section 1, and shows that it is possible to cross the sector boundary while totally missing any HDR.

The region around the HCS, filled in this case with a flux rope, is, as usual, associated with a low proton temperature. Contrary to the flux ropes in Figures 2 and 4, this flux rope exhibits a drop-out of suprathermal electron strahl, indicating that it is totally disconnected from the coronal magnetic field. In the next section, we will investigate further the magnetic connection of blobs and flux ropes via a survey of their properties measured during several HCS crossings in solar cycle 24.

## *3. Survey of the in situ properties of sector boundaries during solar cycle 24*

Depending on how magnetic reconnection takes place in the solar corona, the flux rope associated with a small transient could either be magnetically disconnected from the solar corona, or connected at one end or both ends in the low corona. The way in which magnetic reconnection takes place during the formation of blobs could have implications for the contribution of blobs to the cyclic variation of the total open solar magnetic flux (e.g. Owens & Lockwood 2006). Reconnection between a pair of open magnetic field lines generates a disconnected plasmoid, and a collapsing loop, that would reduce the amount of open magnetic flux. On the other hand, reconnection between closed coronal loops, leading to the expansion of a twisted magnetic loop, could act to open magnetic flux into the heliosphere. In situ measurements of blobs thus provides valuable information about the reconnection process that led to their formation.

We carried out a survey of sector boundary crossings detected in situ from the launch of the STEREO mission until the loss of contact of STEREO-B (2006 - 2014). We focused on boundaries associated with source regions where the coronal neutral line at the Sun underwent strong latitudinal excursions. To identify such neutral lines, we examined PFSS reconstructions of the source surface magnetic field. We consider a neutral line as "highly tilted" when the angle between the neutral line and the equatorial plane exceeds 45° at all latitudes from 30° south to 30° north. The synoptic maps used in this analysis were based on observations by the Wilcox Solar Observatory. In addition, we used the propagation tool (Rouillard et al. 2017) to predict the arrival, at STEREO-A/B and Wind, of the plasma released by this neutral line and looked for the sector boundary crossing closest to the computed arrival time. This provided a preliminary list of 57 crossings. Note that, as a consequence of the high tilt of these neutral lines and associated HCSs, and the near-constant latitude of the selected spacecraft, there are no repeated crossings of the sector boundary, and hence crossings can be given a precise time of passage. We investigate the in situ signatures of horizontal neutral lines in section 4.

From the preliminary list, we remove all cases where a Coronal Mass Ejection (CME) may have been detected in situ during the sector boundary crossing. We consider that – when a CME observed by a coronagraph or heliospheric imager is directed within ±90° of an in situ spacecraft, and when that CME's estimated arrival time is within less than one day before or after the sector boundary crossing – this crossing could be contaminated by a CME passage and must be ignored. CME arrival times were calculated assuming a fixed-phi (i.e. point source) geometry (e. g. Rouillard et al. 2010a) using the propagation tool. The direction of propagation was calculated using simple triangulation, by combining observations from the three coronagraph packages on board the Solar and Heliospheric Observatory (SOHO; Domingo et al. 1995), STEREO-A and STEREO-B when available. When one of these three coronagraphs is not available, the triangulation was only made with two coronagraphs.

The crossings, during the interval extending from 2006 to 2014, that fulfil the above criteria are listed in Table 1. All of the examples shown in Figures 2 to 6 can be found in this table. The first two columns display the detecting in situ spacecraft and the date and time of the observed sector boundary crossing. The third column details the type of trajectory inferred from our analysis of the in situ data, and the nomenclature pertaining to that crossing as introduced in Section 1. The fourth column details whether the HDR, when measured, is composed of multiple peaks separated by small-scale flux ropes, similar to Figure 4 ["YES" denotes that such substructures, with two or three smaller-scale density peaks, were found inside the HDR, "NO" indicates a HDR with a single density peak, and "-" designates cases in which no HDR is detected (scenarios "HCS" and "flux rope")]. Finally, a fifth column is reserved for comments. Note that the dates of all events in Table 1 correspond to the maximum of solar cycle 24. This is because highly-tilted coronal current sheets are mostly observed during solar maximum. During solar minimum, the current sheet is usually oriented in a more east-west direction.

Within this list of crossings there are 4 of type "flux rope + HDR", 4 of type "HDR + flux rope", 2 crossings of a flux rope only, 5 crossings of a HDR only and 1 classical HCS crossing. The remaining two cases are difficult to determine due to the occurrence of well-developed CIRs. As expected, the probability of traversing a flux rope + blob crossing is similar to the probability of traversing blob + flux rope crossing. The probability of only crossing a HDR is larger than the probability of only crossing a flux rope. According to SD17A in their sudy of Carrington rotation 2137, the time taken for a HDR to cross an HI1 pixel is of order 9 hours. The time between the passages of consecutive HDR over a pixel (corresponding to the duration of the intervening dark patch) was found to be on the order of 10h30. This is not consistent with the fact that we detect more HDR crossings than flux rope crossings. However, as was already pointed out by SD17A, the size and periodicity of HDR may vary from interval to interval and, of course, the present statistical study is based on a relatively limited number of events. The probability of crossing a 'clean' HCS is particularly small; only 1 out of the 16 crossings corresponds to such a scenario.

Nine out of the twelve HDRs clearly sampled are associated with a strahl drop-out. On occasion, this drop-out lasts for only a fraction of the blob duration. We can thus conclude that blobs typically carry disconnected magnetic field lines. On the other hand, only 2 out of 11 flux ropes are associated with a strahl drop-out. The remaining 9 flux ropes are associated with field lines connected to the corona, with 7 attached to one hemisphere of the Sun (unidirectional strahl), and 2 attached to both hemispheres (bidirectional strahl). The present analysis thus shows that flux ropes are, in most cases, magnetically connected to at least one hemisphere of the Sun.

Out of the 12 HDRs detected, 5 are sampled alone, without any flux rope. All four are characterized by a strahl electron drop-out. We expect that the spacecraft crossed these 4 blobs near their center, across their maximum radial extents. The substructure of all these four blobs consisted of two or three smaller-scale density enhancements, separated by magnetic fields of low variance and with smoothly-varying components, reminiscent of smaller-scale flux ropes (e.g. Rouillard et al. 2011), similar to Figure 4. Out of the 8 HDRs that were likely sampled near their edges (trajectories 1 and 4), 3 presented similar substructure but with two density peaks. We therefore conclude that HDRs comprise complex structures, consisting of density enhancements and flux ropes at smaller scales.

## *4. Long-lasting HCS crossings in the inner heliosphere*

The analysis of the in situ measurements of crossings of north-south oriented sector boundaries, presented in previous sections, is based on the model sketched in Figure 1. This model was derived by SD17A, from remote-sensing observations made by the inner Heliospheric Imager on the STEREO-A spacecraft (HI1-A) during the passage of a highly-tilted neutral line through the STEREO-A plane of sky during May and June 2013. If this model is correct, a spacecraft crossing a HCS with a significant east-west extent in the ecliptic plane will detect a succession of blobs, followed by flux ropes, with a 20 or so hour periodicity.

The periods of very slow solar wind presented by Sanchez-Diaz et al. (2016) correspond to long-lasting, east-west crossings of near-east-west oriented HCSs by the Helios spacecraft. During these passages, Helios remained close to the HCS for several weeks. During such an extended period, many blobs would be expected to emerge from the solar corona. Figure 1c illustrates a sketch of the distribution of HDRs and flux ropes along a highly-tilted HCS in a meridional plane. For the case of a low latitude HCS oriented in an east-west direction, Figure 1c would, conversely, represent the distribution of small transients in the equatorial plane. The high-density bands sketched in Figure 1c outflow perpetually, at a rate of approximately one every 20 hours. Therefore, if this model is correct, a spacecraft orbiting the Sun at constant latitude near the sector boundary and remaining close to the sector boundary for a number of days, as Helios did during the time periods of very slow solar wind presented by Sanchez-Diaz et al. (2016), should be crossed by a succession of blobs followed by intervening flux ropes.

In the cases presented here, at 1 AU, the blobs were characterized by a HDR and low magnetic fields, i.e., high $\beta$. The flux ropes that followed the blobs were characterized by a strong, often rotating, magnetic field and low density, i.e., low $\beta$. The plasma sheet, containing this succession of blobs separated by flux ropes, was systematically associated with low rather proton temperatures.

Figure 7 shows one of these long-lasting sector boundaries observed by Helios 1 at 0.5 AU (top panels) and by Helios 2 at 0.35 AU (bottom panels). The blobs and flux ropes associated with sector boundaries, presented in Sections 2 and 3 of the current paper, were associated with low temperatures in a similar way to that evident during the entire period of the Heliospheric Plasma Sheet (HPS) delimited in Figure 7 by solid vertical red lines. The HPS extends from 01:00 UT on 1977 October 23 to 21:00 UT on November 03 in the case of Helios 1, and from 09:00 UT on 1977 October 29 to 08:00 UT on November 03 for Helios 2.

The periods of high density, low magnetic field and high $\beta$ are delimited with black dashed lines in Figure 7. We identify these periods, corresponding to the passage of blobs, with the letter B. The flux ropes, observed most clearly at Helios 1 at 0.35 AU, are marked with the letters FR. The HPS is thus typically filled with a series of blobs separated by flux ropes. Flux ropes at Helios 2 are less obvious, although smooth B field regions are clearly present between several of the blobs obseved. Blobs are detected every 18 hours by Helios 2 and every 24 hours by Helios 1. These periodicities are very similar to the periodicity of blob release found by SD17A (19.5 hours). The blobs last around 10 hours at 0.35 AU and 14 hours at 0.5 AU, translating to average radial sizes of 13 and 18 $R_s$, respectively. This is similar to the radial extent of the bright regions of 12 $R_s$, derived by SD17A in the case of a highly-tilted neutral line.

## *5. Discussion*

We have taken advantage of the 3D spatial and temporal distribution of blobs in the heliosphere derived by SD17A in order to identify the signatures of blobs in the in situ measurements of sector boundary crossings. We have also identified their magnetic connection to the solar corona. Our novel approach for comparing remote-sensing observations with in situ measurements is very different from methods employed in previous studies (e. g. by Rouillard et al. 2010a,b and Plotnikov et al. 2016). Instead of tracking the trajectories of individual blobs out to their predicted impact at a spacecraft, we derived the likely 3D distribution of such transients in the heliosphere, using the results of coronal and heliospheric imagery (SD17A), and evaluated the possible trajectories of the spacecraft through such a distribution. This 3D picture of the space-time distribution of blobs can be tested with a wide range of in situ measurements, even during periods when there are no exploitable remote-sensing heliospheric observations. This includes periods when there are no well-developed CIRs.

Among the possible trajectories of a spacecraft through a 3D distribution of blobs (scenario in Section 2), as depicted in Figure 1, three would sample a blob, either totally or partially. Blobs are associated with weaker magnetic fields than the surrounding solar wind, associated with dropouts of strahl electrons. Both signatures are suggestive of magnetic reconnection taking place in the solar corona, and producing a reconnection region that is either partially or totally disconnected from the Sun, in a fashion similar to that reported for larger-scale CMEs (e.g., Shodhan et al., 2000; Riley et al. 2004; Lavraud et al., 2011). On the other hand, most intervening flux ropes show unidirectional suprathermal electron strahl, but a few are associated with bidirectional electrons or strahl electron drop-out. This indicates that the flux ropes that separate blobs are usually attached to one hemisphere of the Sun but in some cases can be connected at both ends or even disconnected from the Sun.

According to these measurements, and the remote-sensing observations presented in SD17A and SD17B, the most likely scenario for the formation of blobs is the one shown in Figure 8a for a plane containing the neutral line and in Figure 8b for a plane perpendicular to the neutral line. For simplicity, Figure 8 is in a reference frame rotating with the Sun, i.e., along the Parker spiral. Reconnection along a large extent of the neutral line produces a series of collapsing closed loops, which are observed in coronal images as raining inflows. Above these collapsing closed loops, a band of complex twisted magnetic field lines, attached to the Sun, expands outwards in the form of flux ropes, similar to the scenario proposed by Gosling et al. (1995) for the onset of CMEs. Between consecutive flux ropes, the reconnection produces HDRs that propagate outwards following the expansion of the flux rope. This reconnection process produces (1) outward-moving flux ropes with enhanced magnetic fields whose main axis is along the direction of the neutral line, (2) intervening broader outflow regions propagating outwards that are observed in white-light images as "blobs" and measured in situ as HDRs with weak magnetic fields, and (3) collapsing loops that are observed in images as raining inflows in the corona.

In this model, succesive reconnection lines (or sequential X-lines) produce the flux ropes, in analogy to one of the models for Flux Transfer Events (FTE) formation at the Earth's magnetopause (Raeder, 2006; Hasegawa et al. 2010). The core of the flux ropes may thus be connected to the corona at either one end, two ends or neither end, depending on the large-scale connectivity of the field lines at the extremity of the reconnection lines. This explanation also bears resemblance to that provided for FTEs at the Earth's magnetopause by Pu et al. [2013]. The in situ measurements reveal, on the other hand, that the field lines associated with blobs, in between the flux ropes, are mostly disconnected from the solar corona. The scenario depicted in Figure 8b also explains these observations. The magnetic field lines around the neutral sheet convect toward the reconnection lines in the direction indicated by the orange arrows. By construction, the outermost field lines (farthest from the symmetry axis of Figure 8) that are reconnecting closest to the Sun engulf the entirety of all other field lines potentially reconnecting at X lines in between the flux ropes. This process means that the HDR or blob regions, where X lines are located, should for the most part be disconnected from the Sun while flux ropes may still be connected to the Sun through their axis, as measured in situ.

The spatial and temporal scales associated with the formation of blobs are corroborated by the measurement of long-lasting sector boundary crossings. When the Helios spacecraft sampled the same sector boundary for several days, they measured alternating blobs and flux ropes. Helios measured one blob and flux rope approximately every day. This is very close to the periodicity of blobs in HI1-A images found by SD17A during solar cycle 24. The radial size of blobs derived from Helios 1 and 2 measurements is 13 and 18 $R_s$ respectively, again very similar to the radial size of blobs observed with HI reported by SD17A (12 $R_s$).

Regarding the internal structure of blobs, the four crossings of a north-south oriented sector boundary that intersect only a blob reveal that the boundary structure, with a succession of blobs separated by flux ropes, is, in fact, repeated at smaller scale inside each blob. This may explain why the signatures of both the front and rear edges of blobs in running-difference J-maps appear to correspond to density increases (SD17A). Some of the blobs partially sampled (flux rope + blob and blob + flux rope type crossings) exhibit such small-scale substructure; the number of enhancements in density measured in these cases tends to be lower than that measured during blob-only crossings. It is very likely that partially sampling blobs in this way fails to reveal their complete structure, showing only two peaks at most, because the spacecraft crosses the edge of the blob rather than their centre.

The typical spatial dimentions and temporal (3 hours) periodicities of the substructure of blobs found here are similar to those of the PDSs found by Viall et al. (2008, 2009) and Kepko et al. (2016) from their examination of in situ measurements. Moreover, these scales are similar to the ones revealed by examination of coronagraph images (Viall & Vourlidas 2015). These results strongly suggest that the PDSs (Viall et al. 2008, 2009, 2010; Viall & Vourlidas 2015; Kepko et al. 2016; Di Matteo et al. 2019) are associated with the substructure of "Sheeley" blobs (Sheeley, 1997). The structure of PDSs, similar to that of blobs but at smaller scales, suggests that they are created through similar mechanisms to those that form the blobs. It is possible that magnetic reconnection could take place within the blobs themselves, giving rise to disconnected plasmoids at the scales of PDSs, or before the formation of blobs, making the later blob enclose several pre-formed PDSs.

In a further study, we aim to better characterize the substructure of blobs using compositional data. Preliminary results of that work support the proposition presented here that HDRs are made up of smaller-scale structures with the typical sizes of PDSs. This includes the HPS in Figure 3, which, on examination of proton density only, could be interpreted as a single large-scale HDR but comprises two clearly-separated structures according to the helium abundance. This may challenge the idea that the HCS could reform after the formation of a HDR (Scenario 2).

# 6. Conclusions and perspectives

Inspection of in situ measurements of the small transients that surround north-south oriented sector boundary crossings complements the remote-sensing studies of blobs presented in SD17A and SD17B.

In SD17B, we showed that the formation of blobs is associated with the simultaneous formation of raining inflows, a likely consequence of magnetic reconnection in the corona. This suggests that the part of the SSW that is sampled as blobs in the white-light images is of transient origin. In SD17A, we quantified the contribution of this transient component to the SSW and found that it could explain one of the two known sources of the wind. In the present paper, we used results of SD17A on the distribution of blobs and their HDRs in the heliosphere to interpret the in situ measurements of small transients at 1 AU. We also revisited some of the in situ measurements of very slow solar wind associated with sector boundary crossings closer to the Sun that were presented by Sanchez-Diaz et al. (2016).

Such inspection of in situ measurements has resulted in a better understanding of (1) the nature of small-scale transients, in particular their origin and their connection to the coronal magnetic field and (2) the structure of blobs and their relationship to smaller-scale transients (Viall et al. 2008, 2009, 2010; Viall & Vourlidas 2015; Kepko et al. 2016). Additionally, the in situ measurements of long-lasting HCS crossings at 0.35 and 0.5 AU allowed us to validate the results of SD17A regarding the spatial and temporal scales for the release of small transients and their distribution in the heliosphere.

In order to explain our measurements, as well as the observations presented in SD17A and SD17B, we propose the following model for the formation of blobs. Reconnection along a large extent of the neutral line produces a series of collapsing closed loops, observed in coronal images as raining inflows. Above these collapsing loops, a band of complicated twisted magnetic field lines attached to the Sun expands outwards in the form of flux ropes. In between consecutive flux ropes, reconnection lines contain HDRs that propagate outwards following the expansion of the flux rope. These dense bands are observed in white-light images as bursts of outpouring "blobs".

One consequence of this model is that a spacecraft crossing the center of a newly-formed flux rope should measure a magnetic field with a stronger component along the direction of the neutral line, while a spacecraft crossing the edge of such a flux rope would measure a rotation of the magnetic field contained in the plane perpendicular to the neutral line. Similarly, the magnetic field of blobs would be mostly contained in the plane perpendicular to the neutral line, but it would remain weak. Due to the compression existing ahead of the CIR, these signatures are not identifiable at 1 AU. The Parker Solar Probe, during its close approaches to the Sun, will provide in situ measurements of blobs near their formation region that will serve to test our predictions.

We finally note that all the measurements shown in this paper were taken during maxima in solar activity. Solar Orbiter, with its hightly-inclined orbits, will cross east-west oriented HCSs, typical of solar minima, in a north-south direction, similar to the east-west crossings of north-south oriented neutral lines presented in this study. This will allow us to repeat this study during the next solar minimum, and eventually enable us to ascertain solar cycle variations in the formation of blobs, and especially the spatial and temporal scales involved in this process.

*Figures*

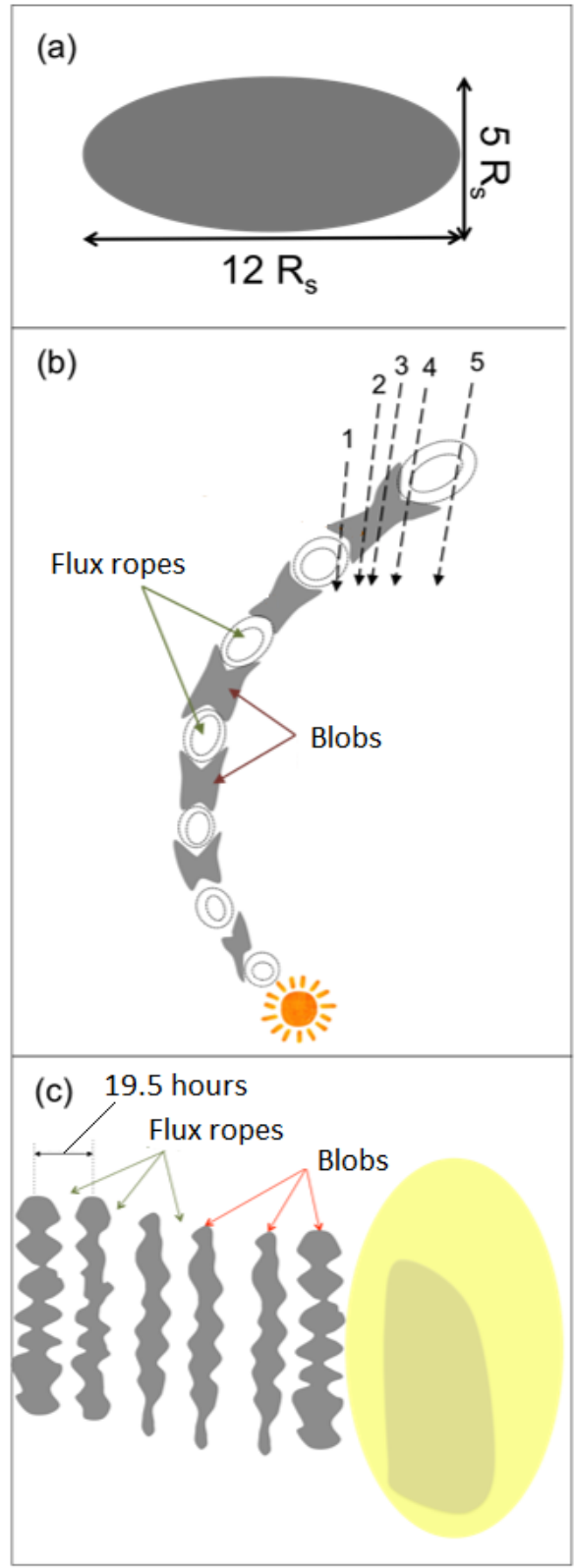

**Figure 1:** (a) Shape and size of a latitudinal cut through the HDR (blobs) associated with a small transient. (b) View from above of the ecliptic plane with blobs, in dark gray, released by an exactly corotating source. The Sun, center bottom, is not to scale but the size of the blobs and the distances involved are. The black dashed elipses indicate the likely location of flux ropes. The black arrows indicate different possible trajectories of a spacecraft throught the pattern of blobs/flux ropes. (c) Latitudinal cut through the heliosphere with its magnetic structure and with blobs in dark grey. The Sun, on the right side, is again not to scale. Adapted from SD17A.

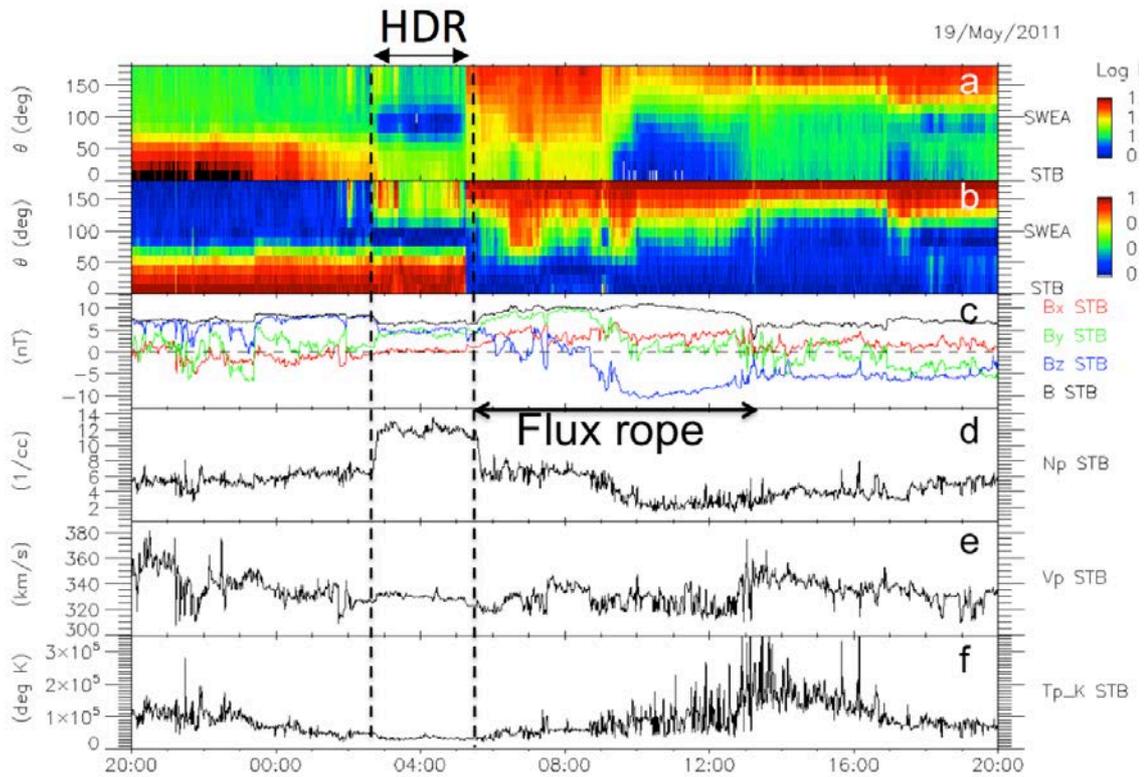

**Figure 2.** In situ measurements of the solar wind by STEREO-B from 20:00 UT on 2011 May 19 to 20:00 UT on May 20. From top to bottom the following quantities are plotted as a function of time: (a) pitch angle distribution of suprathermal electrons at ~250 eV, (b) normalized pitch angle distribution of suprathermal electrons at ~250 eV, (c) magnetic field magntiude and components, (d) proton density, (e) proton bulk speed and (f) proton temperature. The vertical black dashed lines, and the arrow above panel a mark the passage of the spacecraft through a HDR; the arrow below panel c indicates the passage of the spacecraft through a flux rope.

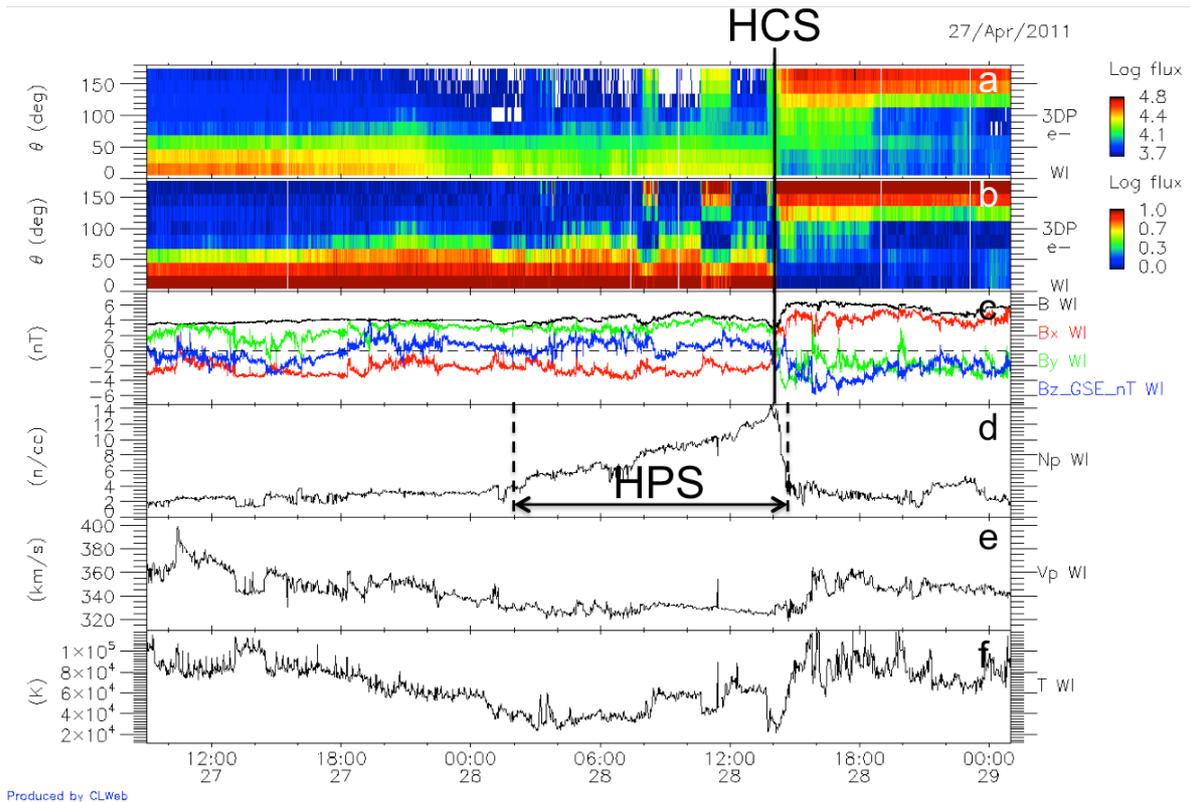

Figure 3. In situ measurements of the solar wind by Wind from 09:00 UT on 2011 April 27 to 00:00 UT on May 29 in the same format as Figure 2. The vertical black dashed lines and the arrow in panel d mark the passage of the spacecraft through the HPS. The black solid line on panels a-c indicates the passage of the HCS.

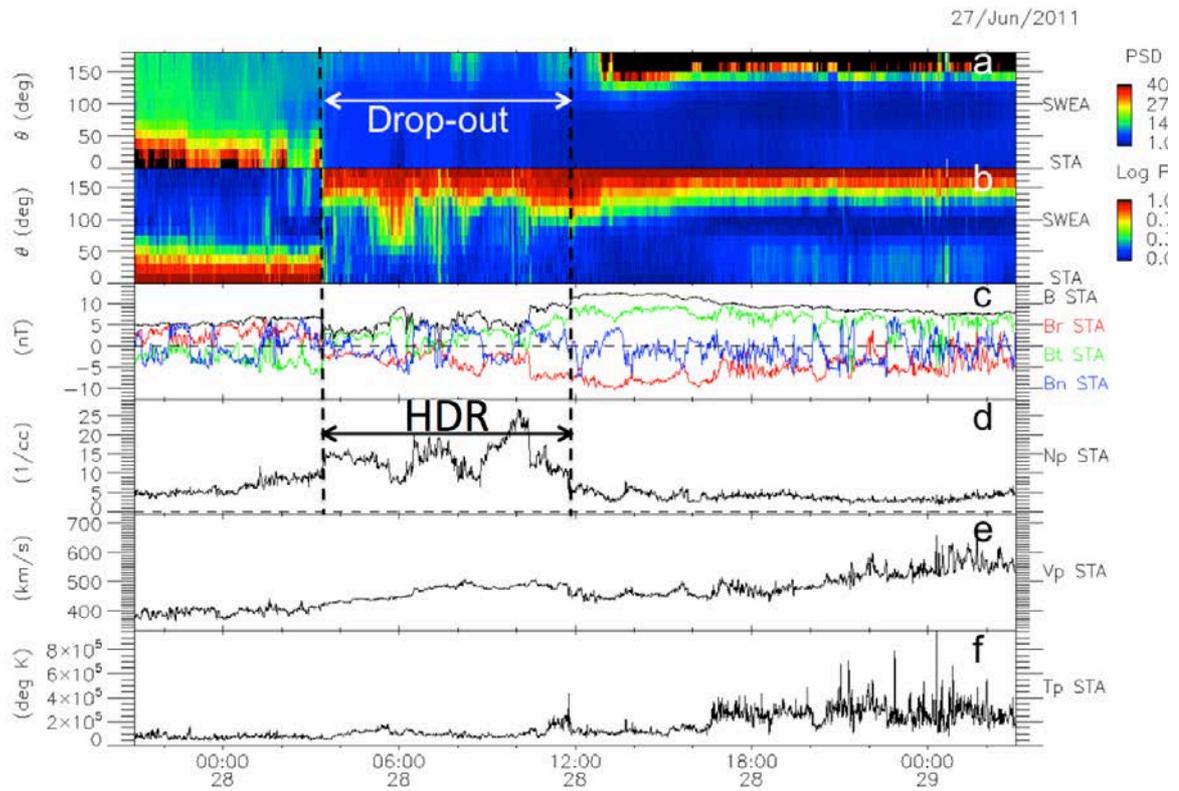

**Figure 4.** In situ measurements of the solar wind by STEREO-A from 09:00 UT on 2011 June 27 to 03:00 UT on June 29 in the same format as Figures 2-3. The vertical black dashed lines and the arrows mark the passage of the spacecraft through a HDR and associated strahl drop-out.

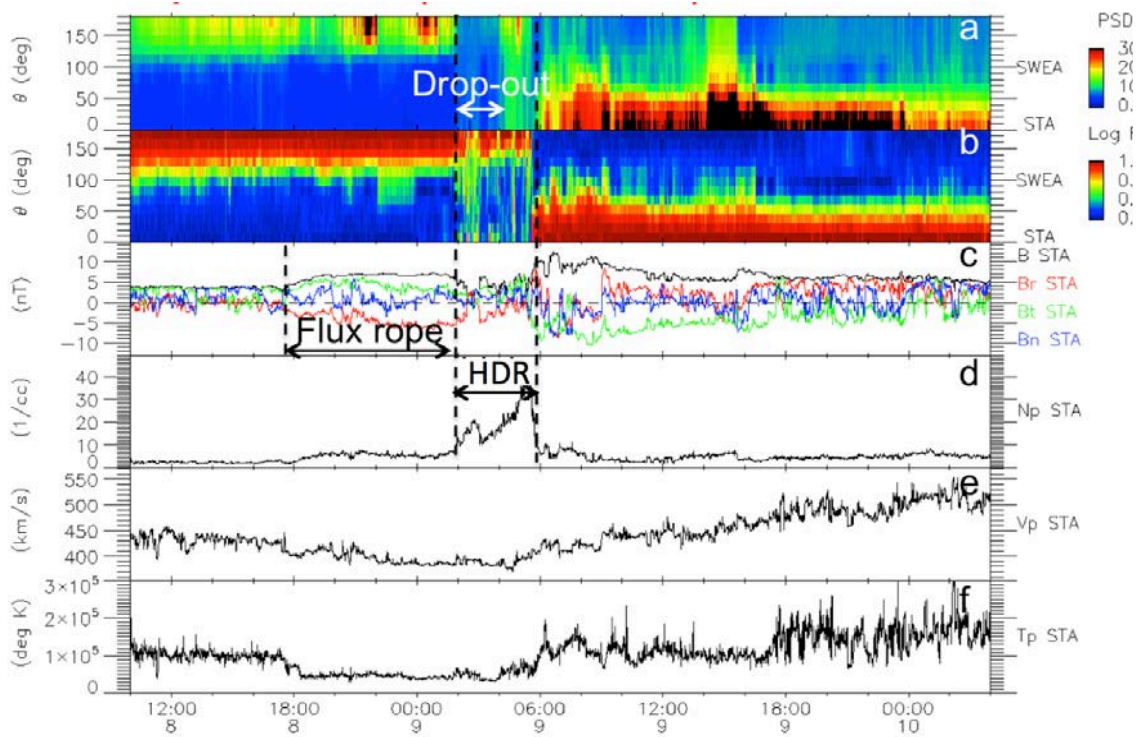

**Figure 5. In situ measurements of the solar wind by STEREO-A from 11:00 UT on 2013 July 08 to 04:00 UT on July 10 in the same format as Figures 2-4. The vertical black dashed lines and arrow in panel c mark the passage of a flux rope. The second pair of vertical black dashed lines and arrow in panel d mark the passage of a HDR. Reproduced from SD17A.**

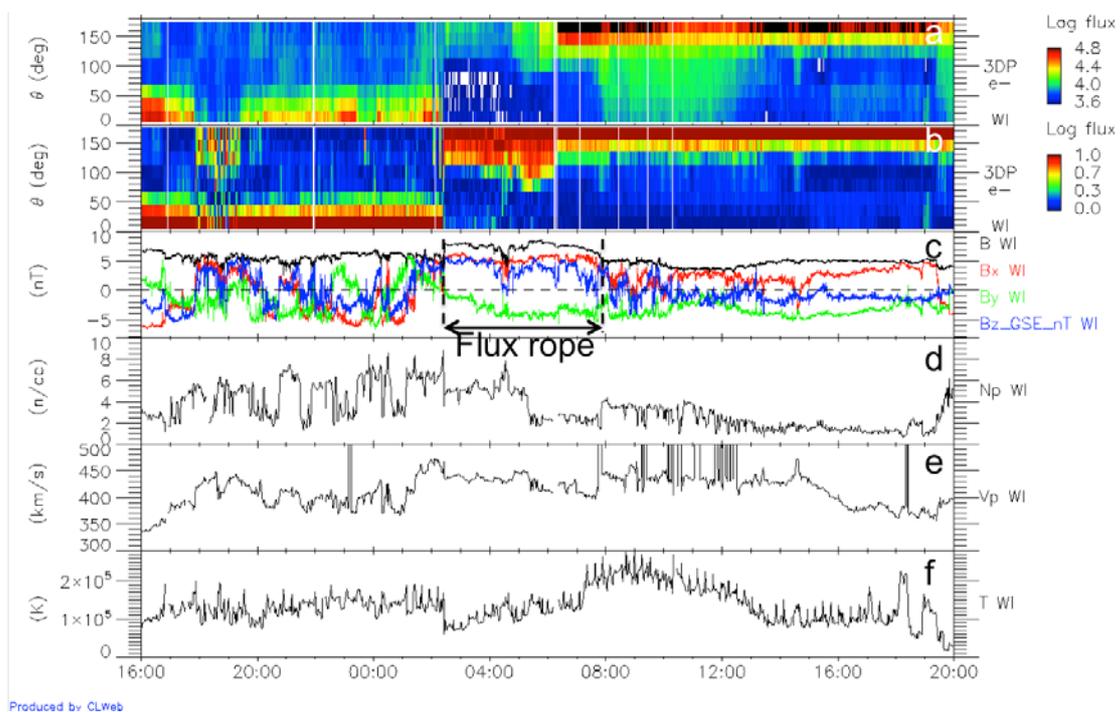

**Figure 6. In situ measurements of the solar wind by Wind from 16:00 UT on 2010 January 30 to 20:00 UT on January 31 in the same format as Figures 2-5. The vertical black dashed lines and arrow in panel c mark the passage of a flux rope.**

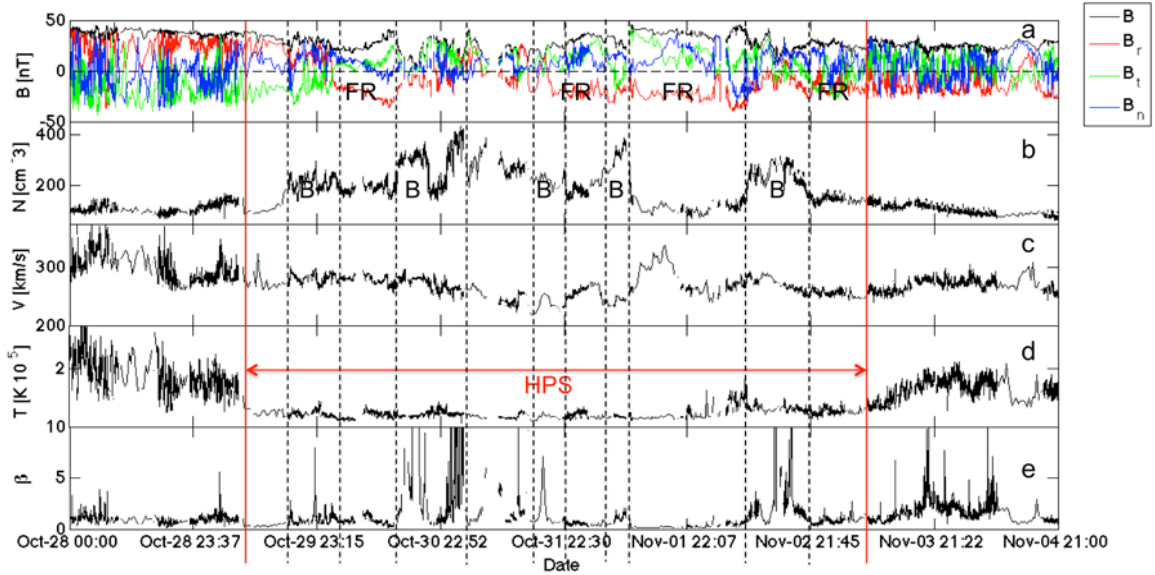

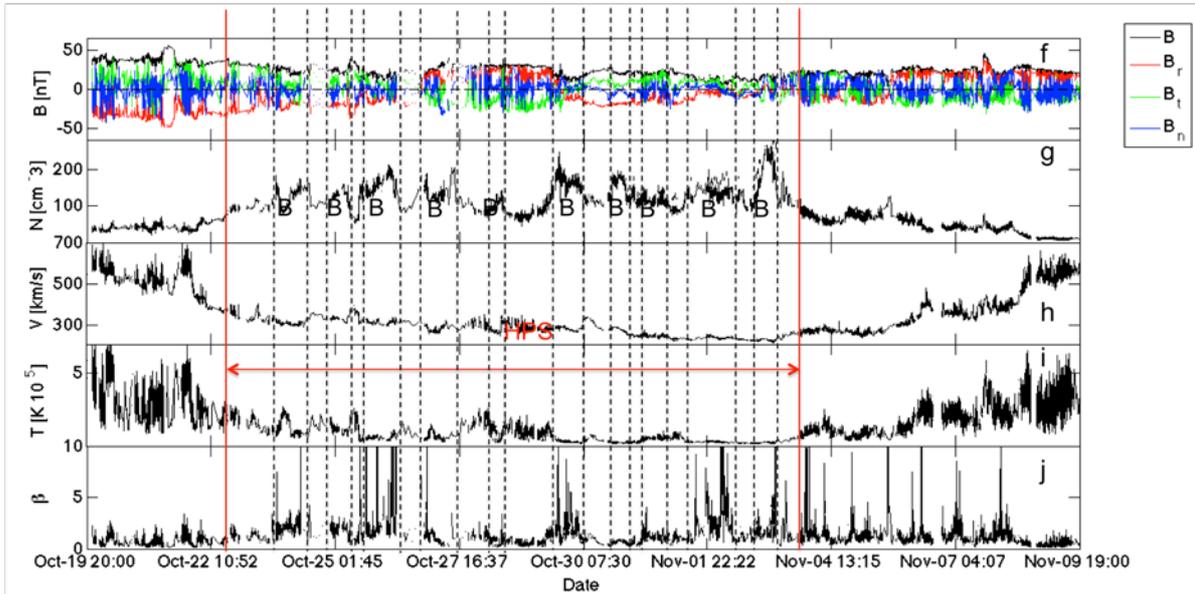

**Figure 7.** In situ measurements from Helios 2 (a-e) and 1 (f-j) of a long-lasting HCS crossing that took place during 1977. Both spacecraft underwent east-west crossings of the same east-west oriented HCS at two different radial heliocentric distances (0.35 AU for Helios 2 and 0.5 AU for Helios 1). For each spacecraft, the quantities shown as a function of time are (from top to bottom): magnetic field magnitude and components (a, f), proton density (b, g), proton bulk speed (c, h), proton temperature (d, i) and β (e, j). This figure covers the same time period as that presented in Figure 1 of Sanchez-Diaz et al. (2016).

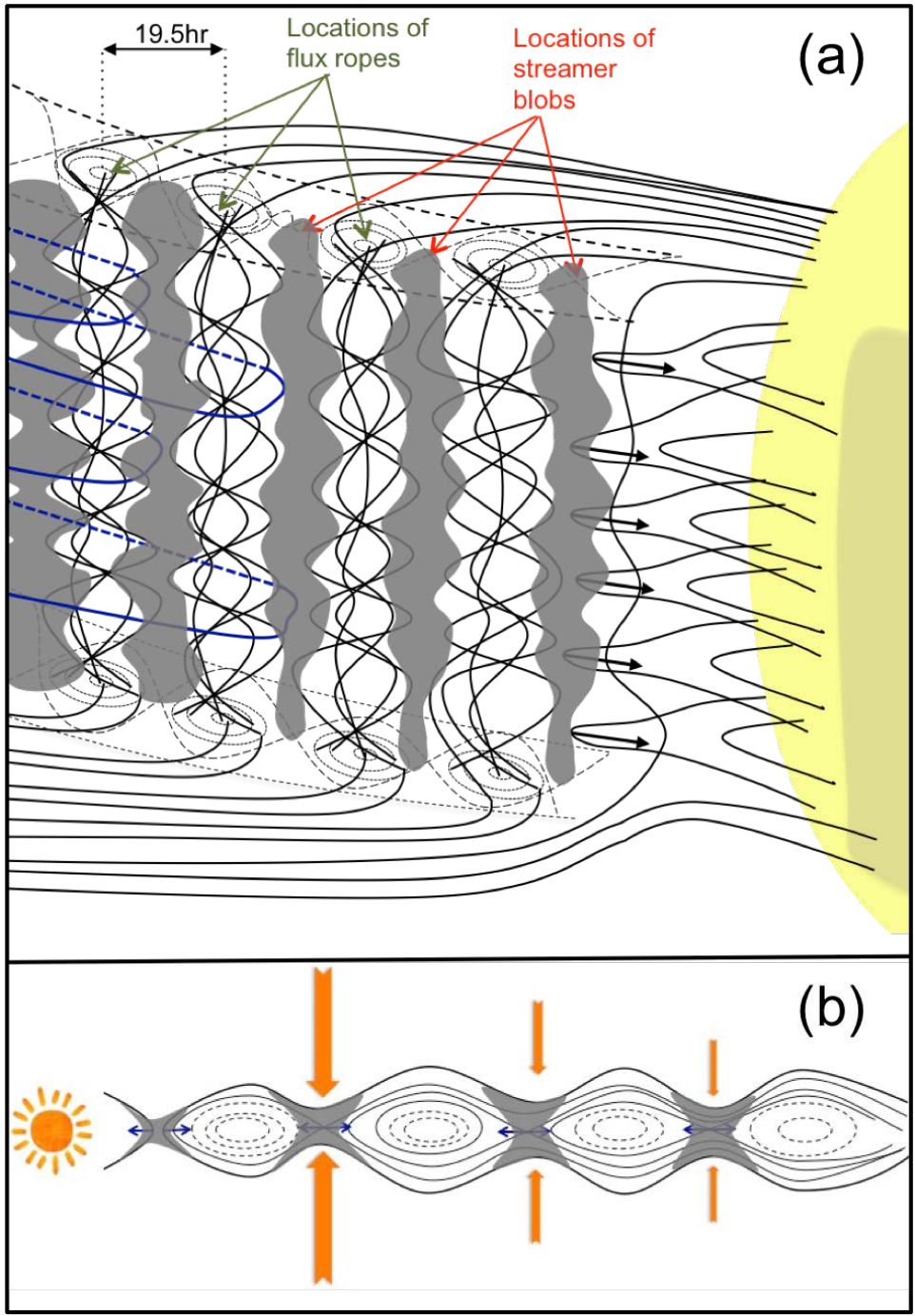

**Figure 8.** Sketch of magnetic reconnection as the origin of blobs in (a) a plane containing the neutral line and (b) a plane perpendicular to the neutral line. The gray areas indicate the location of the HDRs or blobs. The black lines represent the magnetic field lines around the HCS. The dashed black lines represent the magnetic field lines of the flux ropes. The orange arrows show the direction of movement of the field lines prior to magnetic reconnection. The blue arrows indicate the direction of the magnetic pressure following magnetic reconnection

| Spacecraft | HCS crossing date (dd/mm/yy)/time | Trajectory | Sub-blob | Comments |
|---|---|---|---|---|
| STEREO-B | 25/01/10 01:00 | FR (D) + blob | NO | |
| WIND | 31/01/10 01:30 | FR (D) | - | |
| STEREO-B | 24/02/10 13:30 | blob (D) + FR | YES | |
| STEREO-A | 03/04/10 19:00 | blob (D, BDE) | YES | Several short intervals of D and BDE |
| STEREO-B | 20/02/11 19:00 | ¿? | - | Large CIR |
| WIND | 28/02/11 21:30 | blob (D) + FR | NO | Several short intervals of D within the blob |
| STEREO-A | 03/03/11 11:45 | FR | - | |
| WIND | 28/04/11 12:00 | HCS | - | |
| STEREO-A | 04/05/11 21:30 | blob (D) | YES | |
| STEREO-B | 20/05/11 05:45 | blob (D) + FR | NO | Slight D. Only a factor 5 difference in flux between inside and outside the blob |
| STEREO-A | 28/06/11 03:45 | blob (D) | YES | |
| WIND | 18/07/11 04:15 | FR+blob | NO | |
| STEREO-B | 04/09/11 19:15 | FR (BDE) + blob? | ¿? | Not clear if a blob or CIR |
| WIND | 07/04/13 14:45 | blob (D) | YES | |
| STEREO-A | 09/07/13 05:45 | FR + blob (D) | YES | |
| STEREO-B | 16/07/13 13:45 | CIR + FR | - | |
| WIND | 25/07/13 14:45 | Blob | YES | |
| STEREO-A | 04/08/13 19:15 | blob (D) + FR | NO | The D starts before the blob and continues into the blob |

**Table 1. List of crossings of highly-tilted HCSs not contaminated by CMEs during solar cycle 24. The first column lists the observing spacecraft. The second column displays the date and time of the HCS crossing. The third column reports on the trajectory of the spacecraft, according to the sketch in Figure 1 and description in the Introduction. When a blob is measured, the fourth column states whether it contains substructure of smaller-scale density enhancements and flux ropes, similar to that of Figure 4. The fifth column lists additional comments. FR denotes flux rope, D electron drop-out, BDE bidirectional electrons and CIR Corotating Interaction Region. D or BDE inbrackets following FR or blob means that the FR or blob was associated with these features of the pitch angle distribution of suprathermal electrons. HCS denotes a clean HCS crossing.**